\documentclass[runningheads]{llncs}
\usepackage[T1]{fontenc}
\usepackage{orcidlink}
\usepackage{natbib}
\usepackage{graphicx}
\usepackage{subcaption} 
\usepackage{hyperref}
\usepackage{multirow}
\usepackage{booktabs}
\everypar\expandafter{\the\everypar\looseness=-1}

\begin{document}
\title{Eco-Aware Graph Neural Networks for Sustainable Recommendations}
\titlerunning{Eco-Aware GNNs for
Sustainable Recommendations}
%

 \author{Antonio Purificato \orcidlink{0009-0009-3933-380X} \and
 Fabrizio Silvestri \orcidlink{0000-0001-7669-9055}}

\institute{Sapienza University of Rome, Rome, Italy \\ \email{\{purificato, fsilvestri\}@diag.uniroma1.it}}

\authorrunning{Purificato et al.}
\maketitle              
\begin{abstract}
Recommender systems play a crucial role in alleviating information overload by providing personalized recommendations tailored to users' preferences and interests. Recently, Graph Neural Networks (GNNs) have emerged as a promising approach for recommender systems, leveraging their ability to effectively capture complex relationships and dependencies between users and items by representing them as nodes in a graph structure.

In this study, we investigate the environmental impact of GNN-based recommender systems, an aspect that has been largely overlooked in the literature. Specifically, we conduct a comprehensive analysis of the carbon emissions associated with training and deploying GNN models for recommendation tasks. We evaluate the energy consumption and carbon footprint of different GNN architectures and configurations, considering factors such as model complexity, training duration, hardware specifications and embedding size.

By addressing the environmental impact of resource-intensive algorithms in recommender systems, this study contributes to the ongoing efforts towards sustainable and responsible artificial intelligence, promoting the development of eco-friendly recommendation technologies that balance performance and environmental considerations. Code is available at:\\ \url{https://github.com/antoniopurificato/gnn_recommendation_and_environment}.

\keywords{Graph Neural Networks \and Recommendation Systems \and Environmental Impact.}
\end{abstract}
\section{Introduction}

Recommender systems (RSs) play a crucial role in mitigating information overload by providing personalized recommendations, benefiting users and service providers across various platforms, including e-commerce (e.g., Tmall, Amazon) \citep{10.1145/3397271.3401431} and social networks (e.g., Gowalla, Facebook) \citep{10.1145/3366423.3380112,10.1109/TKDE.2016.2569096}. Different approaches have been proposed, ranging from collaborative filtering techniques that leverage user-item interactions to content-based methods that analyze item features \citep{8506344} to sequential recommendation, aiming to capture the sequential patterns in user behavior and provide recommendations based on the current context or session \citep{10.1007/978-3-031-56060-6_14,bacciu2023integrating}. 

In addition to sequential recommendation, Graph Neural Networks (GNNs) have emerged as a promising approach for recommender systems \citep{10.1145/3626772.3657716,10.1145/3604915.3608804}. GNNs can effectively capture the complex relationships and dependencies between users and items by representing them as nodes in a graph structure. By propagating and aggregating information along the edges of the graph, GNNs can learn rich representations that encode high-order connectivity patterns, leading to improved recommendation performance \citep{10.1145/3535101}. Furthermore, GNNs can naturally incorporate various types of auxiliary information, such as user profiles, item attributes, and social connections, into the graph structure, enabling the exploitation of heterogeneous data sources for more accurate recommendations \citep{purificato2024sheaf4recsheafneuralnetworks}. Nowadays, resource-intensive algorithms have become prevalent in modern recommender systems, resulting in higher energy consumption for recommendation experiments \citep{betello2024reproducible}. However, despite some studies having been carried out in regarding the environmental impact of SRSs \citep{betello2024reproducible,10.1145/3604915.3608840}, only one work presented an analysis of the computational consumption of GNN-based RSs \cite{10.1145/3604915.3608840}.

In this work, we analyse the environmental impact of GNN-based RSs experiments by faithfully replicating representative experimental pipelines. Through a comprehensive comparative analysis, we shed light on the carbon emissions attributable to the training and deployment of GNN-based RSs. Our study serves as a clarion call for sustainability, underscoring the need to reconcile the pursuit of technological advancements with environmental consciousness within the realm of RSs. This study aims to answer to the following research questions:
\begin{itemize}
    \item \textbf{RQ1}: Which model performs the best, and what are the trade-offs in terms of resource consumption?
    \item \textbf{RQ2}: How does the embedding size of the GNN affect the environmental impact of the results?
\end{itemize}

\section{Related Work}
In the Glasgow Agreement \citep{hunter2021glasgow}, participating nations committed to reducing CO\textsubscript{2} emissions, underscoring the urgent need for environmental action. This commitment is particularly relevant to our field, as the environmental impact of GPU training in machine learning is significant. The energy consumption associated with these computational processes contributes significantly to CO\textsubscript{2} emissions, exacerbating climate change \citep{patterson2021carbon}. It is therefore our responsibility to raise awareness of this issue.

Although the environmental impact of deep learning algorithms has been investigated in certain domains, such as Natural Language Processing \citep{10.1145/3442188.3445922,wang2023energy} and Information Retrieval \citep{10.1145/3477495.3531766}, there is a dearth of research examining the environmental footprint of Recommender Systems.

\looseness -1 \cite{10.1145/3604915.3608840} benchmark several state-of-the-art recommendation algorithms
in terms of both recommendation performance and carbon emissions and analyze the trade-off between energy consumption, carbon emissions and the predictive accuracy of recommendation
algorithms. A difference with respect to our approach is that they do not study the impact of some important factors on the performance and on the emissions, such as the embedding size of each model.

\cite{betello2024reproducible} provide a code resource and a robust framework for developing RSs and establishing a foundation for consistent and reproducible experimentation. They also study how the number of parameters influences the CO\textsubscript{2} consumption of the proposed algorithms. Differently from our work, they do not consider GNN-based approaches, but only sequential recommendation algorithms. While their research aims to provide a reproducibility analysis with an associated framework, the objective of our work is to highlight the significance of implementing environmentally sustainable solutions for recommendation tasks.

In the next Section we present our methodological approach to assess the environmental impact of recommender systems.

\section{Method}

\subsection{Calculating CO\textsubscript{2} Emissions}
In this study, we use CodeCarbon\footnote{\url{https://codecarbon.io}} \citep{codecarbon}, a tool designed to track the power consumption of both CPUs and GPUs. This allows us to measure carbon dioxide equivalent (CO\textsubscript{2}-eq), a widely accepted standard used by numerous organisations and governments to monitor emissions of various greenhouse gases \citep{kim2009measurement}. CO\textsubscript{2}-eq facilitates the comparison of greenhouse gas emissions by converting quantities of different gases into an equivalent amount of CO\textsubscript{2}, based on their respective global warming potentials. By using CodeCarbon, we can accurately assess the environmental impact of our training processes, in line with our commitment to sustainability and responsible research practices in machine learning.

Our decision to focus specifically on CO$_2$-eq emissions is motivated by several factors. First, CO$_2$-eq is a widely recognized and accepted metric for quantifying the combined impact of various greenhouse gases on global warming, allowing us to provide a standardized and comparable measure of the overall environmental impact \cite{betello2024reproducible,10.1145/3604915.3608840}. Second, CO$_2$ emissions are particularly relevant in the context of energy-intensive machine learning training processes, which typically rely on electricity generated from fossil fuel sources. Tracking CO$_2$-eq emissions allows us to directly assess the carbon footprint associated with the energy consumption of our experiments.

\subsection{Models}
The Neural Graph Collaborative Filtering (NGCF) model \citep{ngcf}  represents user-item interactions as a bipartite graph and learns user and item embeddings by propagating them on the graph through message passing layers. The key idea is to capture high-order connectivity patterns by recursively propagating embeddings over the graph using graph convolutional layers. NGCF is trained end-to-end to minimize the difference between predicted and actual user-item interactions. It exploits high-order connectivity, learning non-linear relationships, and providing inductive capabilities for new users or items.

LightGCN \citep{lightgcn} learns user and item embeddings by propagating them on the user-item interaction graph through a series of graph convolutional layers. Unlike NGCF, LightGCN removes the feature transformation and nonlinear activation layers, making it a linear model. The key idea is to simplify the GCN architecture while preserving the ability to capture high-order connectivity patterns between users and items. LightGCN is trained end-to-end to optimize the BCE between the predicted and actual user-item interactions. It has demonstrated competitive performance while being computationally efficient.

SimGCL \citep{yu2022graph} add random uniform noise
to hidden representations for augmentations, resulting in more
uniform node representations that mitigate the popularity bias. Adjusting the noise magnitude can improve regulation of representation uniformity, leading to advantages on recommendation accuracy and model training efficiency. 

In LightGCL \citep{lightgcl}, the graph augmentation is guided by singular value decomposition (SVD) to not only distill the useful information of user-item interactions but also inject the global collaborative context
into the representation alignment of contrastive learning. Instead of generating two handcrafted
augmented views, important semantic of user-item interactions can be preserved with their paradigm. This enables self-augmented representations to be reflective of both user-specific preferences and cross-user global dependencies.

\subsection{Experimental Pipeline}
To evaluate the impact of different embedding sizes on the model's performance and computational requirements,  experiments with embedding sizes of 32, 64, 128, and 256 are conducted. The results of these experiments are presented in Section \ref{sec:results}. Prior to commencing the training process, we initialize a CodeCarbon tracker to monitor the carbon emissions associated with the training process. Additionally, we utilize the DeepSpeed\footnote{\url{https://www.deepspeed.ai}} library to compute the number of floating-point operations (FLOPs) required for each model configuration.

During the training process, the CodeCarbon library is used to log the power consumption every 30 seconds, allowing us to track the energy consumption in real-time. Upon completion of the training, we compute the total carbon emissions and various performance metrics, as described in Section \ref{sec:exps}. 

In the next Section we will presents the experimental setup and describe the different metrics and datasets used in the experiments.

\section{Experiments}
\label{sec:exps}

\subsection{Datasets}
Our analyses encompass a collection of datasets, ensuring comprehensive and robust insights. By incorporating datasets with diverse characteristics, such as varying user and item counts, we aim to unravel the intricate interplay between these factors and our findings. This approach enables us to capture a view of real-world scenarios, thereby fortifying the applicability of our conclusions across a broad spectrum of contexts. All the statistics of these datasets are presented in Table \ref{tab:dataset_info}.

\begin{itemize}
    \item MovieLens\footnote{\url{https://grouplens.org/datasets/movielens}}: The MovieLens dataset \citep{10.1145/2827872} is widely recognized as a benchmark for evaluating recommendation algorithms. We utilize MovieLens 1M (ML-1M).
    \item Amazon: These datasets consist of product reviews collected from Amazon.com \citep{10.1145/2766462.2767755}. The data are organized into distinct datasets based on Amazon's primary product categories. For our study, we focus on the ``Beauty'' category (Beauty).
    \item DianPing: This dataset contains the user reviews as well as the detailed business meta data information crawled from a famous Chinese online review website\footnote{\url{DianPing.com}}.
\end{itemize}

Our data preprocessing pipeline adheres to well-established practices in the field. We adopt an implicit approach, treating all interactions as binary events without considering rating values, as done in \citep{kang2018self,sun2019bert4rec}.

For dataset partitioning, we employ a widely-used strategy in sequential recommendation tasks \citep{sun2019bert4rec,kang2018self}. The most recent interaction for each user is held out for testing, while the second-to-last interaction is reserved for validation. The remaining interactions constitute the training set, providing a chronological sequence of user behavior.

\begin{table*}[!ht]
\centering
    \caption{Dataset statistics after pre-processing. Density and sparsity are percentage values.}
  \begin{tabular}{l||ccc|cc}
    \toprule
Dataset name & Users & Items & Interactions &  Density & Sparsity \\ 
    \midrule
Amazon Beauty & 1,210,271	& 249,274 & 2,023,070 & 0.001 & 99.999 \\
MovieLens 1M &  6,040 & 3,952 & 999,611 & 4.189 & 95.810\\
DianPing & 542,706	& 243,247	& 4,422,473 & 0.003 & 99.997\\
\bottomrule
  \end{tabular}
\label{tab:dataset_info}
\end{table*}

\subsection{Metrics}
To evaluate the performance of sequential recommendation algorithms, we employed four widely adopted metrics commonly used in Information Retrieval (IR) \citep{kang2018self,purificato2024sheaf4recsheafneuralnetworks}: Precision, Recall, Normalized Discounted Cumulative Gain (NDCG), and Hit Ratio (HIT). These metrics provide a comprehensive assessment of the recommendation system's ability to identify relevant items and rank them effectively.

\begin{itemize}
    \item Precision: This metric calculates the proportion of correctly identified relevant items among the recommended items. It measures the system's ability to avoid irrelevant recommendations.
    \item Recall: It quantifies the fraction of correctly identified relevant items among the recommendations relative to the total number of relevant items in the dataset. This metric evaluates the system's capability to retrieve as many relevant items as possible.
    \item Normalized Discounted Cumulative Gain (NDCG): This metric evaluates the performance of a ranking system by considering the position of relevant items in the ranked list. It assigns higher scores to relevant items ranked higher, as they are typically where a user's attention is focused. NDCG captures the importance of ranking relevant items at the top of the recommendation list.
    \item Hit Ratio (HIT): Is a key metric in recommendation systems that measures whether relevant items appear within the top K positions of a model's recommendation list. For each user, if at least one relevant item is included in the top K recommendations, it counts as a "hit." The HIT@K score is then calculated as the proportion of users for whom the model successfully includes at least one relevant item within the top K.
    \item Emissions: Represents the CO$_2$-eq (measured in Kg) required for training a single model and is the sum of the single CO$_2$-eq emissions over each epoch.
\end{itemize}
By employing these four metrics, we can comprehensively assess the recommendation system's ability to identify relevant items, rank them effectively, and provide high-quality recommendations tailored to the user's preferences and interests.

\subsection{Reproducibility}
In order to facilitate a rigorous and unbiased comparison, a standardized experimental setup was adopted for all models. The training regime consisted of 400 epochs, with batch sizes of 2048 and 4096 for the training and validation stages, respectively. The optimization was carried out using the Adam algorithm, with a learning rate fixed at 0.001. Furthermore, to mitigate the effects of random initialization and promote reproducibility, an identical seed was employed across all experiments. In order to see how all the experiments evolve over time, no early stopping procedures were applied.

\subsection{Hardware}
All experiments were performed on a single NVIDIA RTX A6000 with 10752 CUDA cores and 48 GB of RAM. The code is written in Python 3 and to train all the models it was used the RecBole library\footnote{\url{https://recbole.io/}} \citep{10.1145/3539618.3591889}.

In the next Section we will present the results of the proposed study, in terms of performance metrics and environmental impact.

\section{Results}
\label{sec:results}

\subsection{RQ1: Which model excels, and at what cost?}
\looseness -1 As shown in Table \ref{table:main}, LightGCN outperforms all competitors across all datasets, with the most significant lead observed on the Beauty dataset, while the gap narrows on the ML-1M dataset. LightGCN's superior performance can be attributed to two key factors: firstly, it is an advancement over NGCF, with its advantages clearly demonstrated in prior experiments \citep{lightgcn}. Secondly, both LightGCL and SimGCL, although promising, involve higher computational complexity, and 400 epochs may not suffice for them to converge to optimal results. In terms of carbon emissions, NGCF remains the most efficient on two out of the three datasets—a somewhat unexpected outcome, given that LightGCN is touted by its authors as being more lightweight than NGCF. The best trade-off in terms of performance-emission will probably remain LightGCN, which is the second-best model in terms of environmental impact on two of the three datasets.

\begin{table*}[!ht]
    \centering
    \resizebox{\linewidth}{!}{
    \begin{tabular}{cc|cccc|cccc|c}
        \toprule
        \textbf{Dataset} & \textbf{Model} & \textbf{P@10} & \textbf{R@10} & \textbf{NDCG@10} & \textbf{HIT@10} & 
        \textbf{P@100} & \textbf{R@100} & \textbf{NDCG@100} & \textbf{HIT@100} & \textbf{Emissions} \\ 
        \midrule
        \multirow{4}{*}{\textbf{Beauty}} & \textbf{LightGCL} & \underline{.0019} & \underline{.0185} & \underline{.0107} & \underline{.0187} 
        & \underline{.0005} & \underline{.0517} & \underline{.0190} & \underline{.0591} & 12.5521 \\ 
         & \textbf{LightGCN} & \textbf{.0022} & \textbf{.0212} & \textbf{.0121} & \textbf{.0217} 
         & \textbf{.0006} & \textbf{.0583} & \textbf{.0195} & \textbf{.0594} & \underline{4.1341} \\ 
         & \textbf{NGCF} & .0017 & .0160 & .0087 & .0164 & 
         .0005 & .0508 & .0156 & .0519 & \textbf{1.1147} \\ 
         & \textbf{SimGCL} & .0004 & .0039 & .0019 & .0039 
         & .0002 & .0227 & .0054 & .0230 & 12.9475 \\ 
         \midrule
        \multirow{4}{*}{\textbf{DianPing}} & \textbf{LightGCL} & \underline{.0044} & \underline{.0239} & \underline{.0221} & \underline{.0499} & 
         \underline{.0018} & \underline{.1673} & \underline{.0393} & \underline{.1734} & \underline{15.011}1 \\  & \textbf{LightGCN} & \textbf{.0053} & \textbf{.0417} & \textbf{.0223} & \textbf{.0518} & 
        \textbf{.0023} & \textbf{.1680} & \textbf{.0482} & \textbf{.2031} & \textbf{13.4835} \\ 
         & \textbf{NGCF} & .0032 & .0235 & .0123 & .0313 & 
         .0017 & .1205 & .0320 & .1523 & 27.3234 \\ 
         & \textbf{SimGCL} & .0040 & .0238 & .0101 & .0474 & 
         .0016 & .1435 & .0347 & .1600 & 21.2422 \\ 
         \midrule
        \multirow{4}{*}{\textbf{ML-1M}} & \textbf{LightGCL} & .1986 & .1591 & .2546 & .7341 & 
        .0761 & .5231 & .3366 & .9538 & \underline{0.1185} \\ 
        & \textbf{LightGCN} & \textbf{.2094} & \textbf{.1731} & \textbf{.2681} & \textbf{.7626} 
        & \textbf{.0808} & \textbf{.5692} & \textbf{.3619} & \textbf{.9664} & 0.2146 \\ 
        & \textbf{NGCF} & \underline{.2001} & \underline{.1623} & \underline{.2549} & \underline{.7373} & 
        \underline{.0786} & \underline{.5492} & \underline{.3470} & \underline{.9621} & \textbf{0.1057} \\ 
        & \textbf{SimGCL} & .1073 & .1188 & .1490 & .6210 & 
        .0438 & .4037 & .2333 & .9386 & 0.1281 \\
        \bottomrule
    \end{tabular}}
    \caption{\looseness-1 Results of the models in terms of Precision@K (P@K), Recall@K (R@K), NDCG@K and HIT@K, with \mbox{K $\mathbf{\in \{10,100\}}$} ($\uparrow$ is better). Also the emissions in terms of CO\textsubscript{2}-eq [kg] are shown ($\downarrow$ is better). \textbf{Bold} denotes the best model for a dataset by the metric in the main group, \underline{underlined} the second best.}
    \label{table:main}
\end{table*}

\subsection{Embedding Size and Environmental Impact}

\begin{figure}[h!]
    \centering
    \begin{subfigure}[b]{.6\textwidth}
        \centering
        \includegraphics[width=\textwidth]{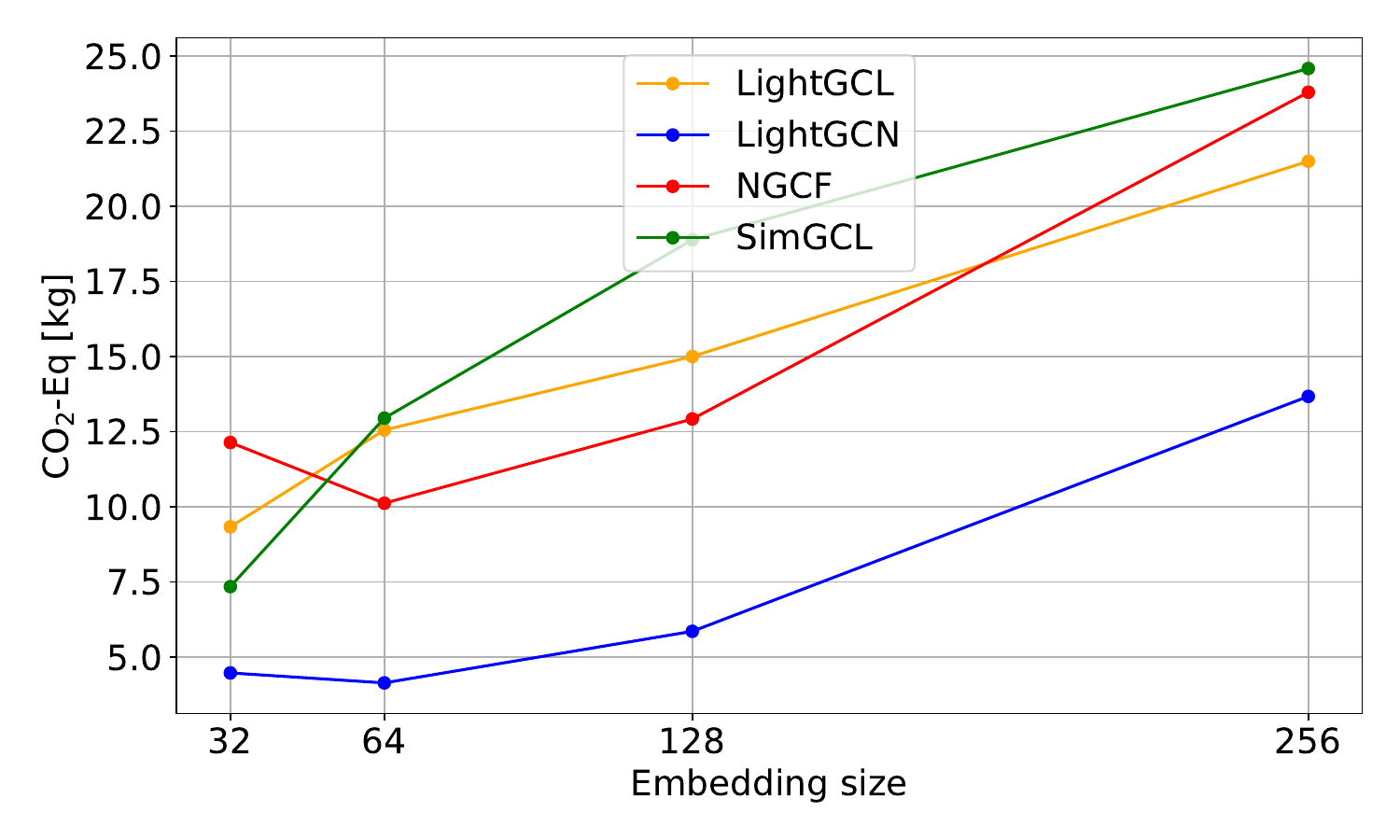} 
        \caption{Beauty}
        \label{fig:sub1}
    \end{subfigure}\\
    \begin{subfigure}[b]{.6\textwidth}
        \centering
        \includegraphics[width=\textwidth]{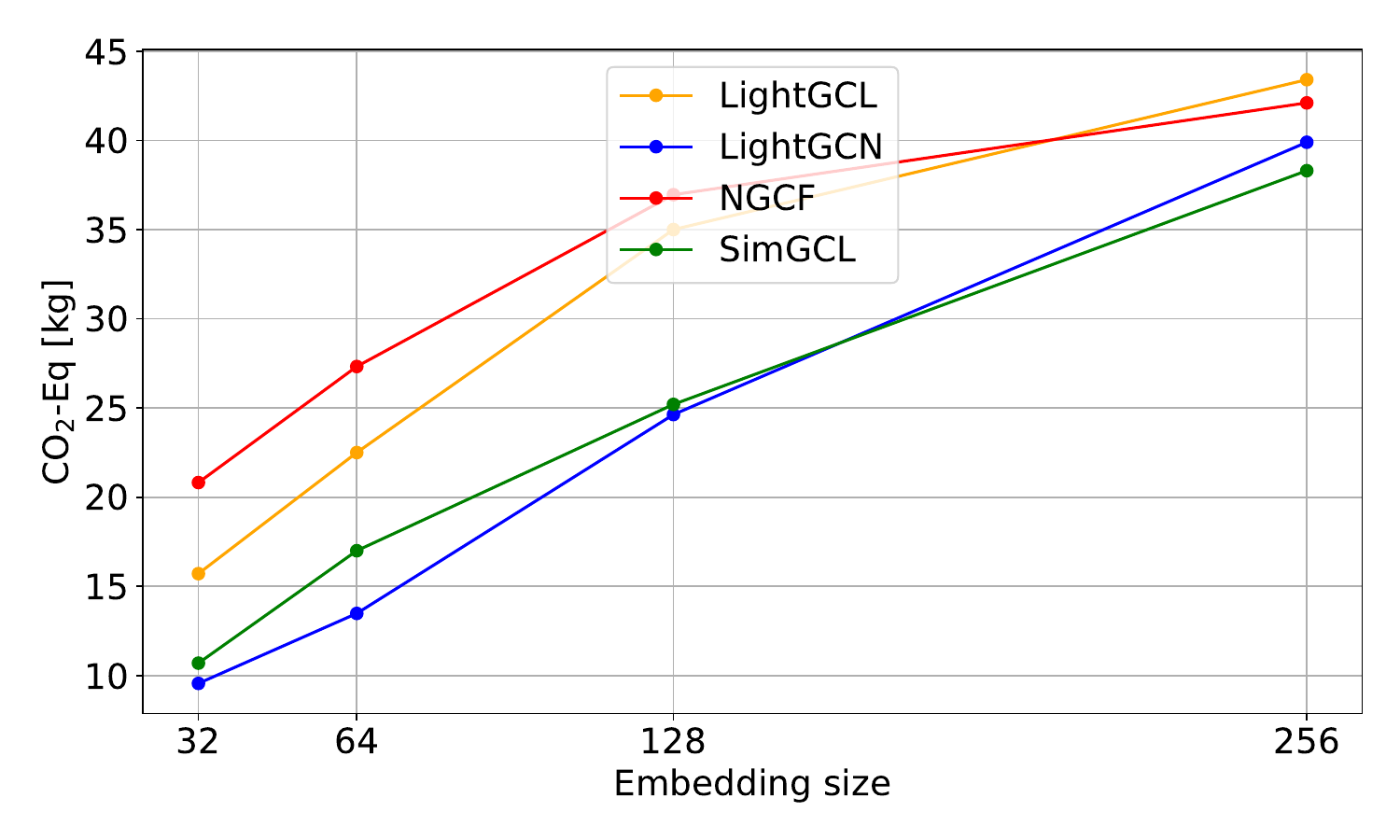} 
        \caption{DianPing}
        \label{fig:sub2}
    \end{subfigure}\\
    \begin{subfigure}[b]{.6\textwidth}
        \centering
        \includegraphics[width=\textwidth]{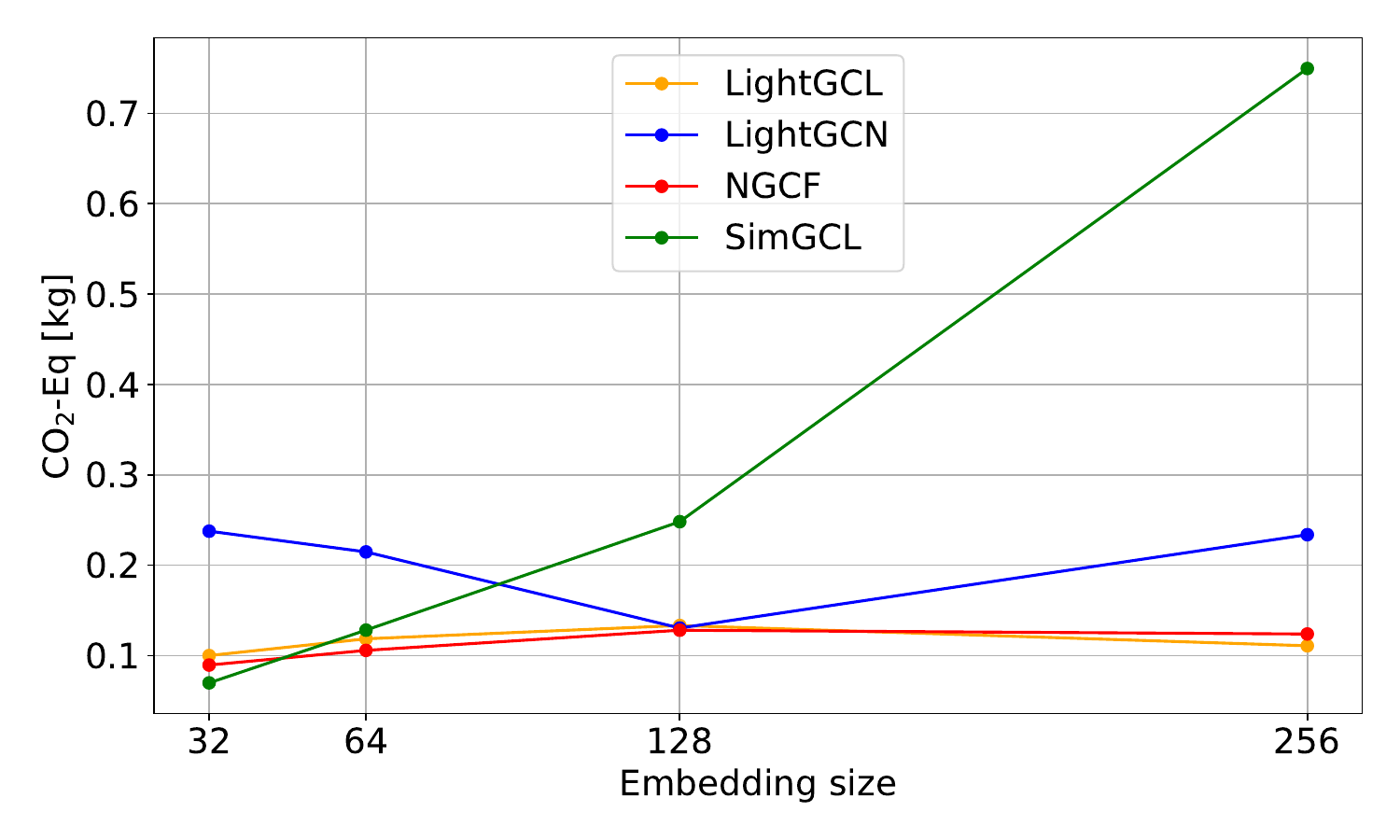} 
        \caption{ML-1M}
        \label{fig:sub3}
    \end{subfigure}
    
    \caption{Difference in terms of environmental impact when changing the embedding size of the GNNs. On the y-axis it can be seen the CO$_2$-eq while on the x-axis the embedding size.}
    \label{fig:main_figure}
\end{figure}

Figure \ref{fig:main_figure} shows that increasing the size of the embeddings generally leads to a higher environmental impact across all models and datasets. This trend is particularly pronounced in the DianPing dataset, which contains a large number of interactions. Interestingly, on the DianPing dataset, NGCF has a higher cost than SimGCL, despite SimGCL being computationally more expensive. However, on the other datasets, SimGCL consistently remains the most computationally demanding model. The same Figure also illustrates how the dataset size affects CO$_2$-eq costs. On the ML-1M dataset, the costs are significantly lower compared to those on the DianPing dataset. This not only depends from the number of users and the number of items, but also from the interactions between users and items. On the DianPing and Beauty datasets, as the embedding size increases, emissions also increase, as expected. It is interesting to note that on the ML-1M dataset, for the NGCF and LightGCL models, emissions related to an embedding size of 256 are higher compared to those with an embedding size of 128. This result requires more detailed investigation in future work, particularly on how each parameter of a model influences the environmental impact of the respective model.

\section{Conclusions}

In this study, we examined the environmental impact of GNN-based recommender systems, an aspect often overlooked in AI research. Our analysis of carbon emissions and energy consumption across different GNN architectures and configurations highlights how model complexity, training duration, hardware specifications, and embedding size affect their environmental footprint.

While GNNs provide significant benefits in capturing complex relationships for recommendation tasks, our findings show that these gains can come with considerable environmental costs, especially when large datasets or extensive embeddings are used.
By emphasizing these trade-offs, our study contributes to the discourse on sustainable AI, encouraging the integration of environmental considerations into the development of recommender systems. Future research should focus on optimizing GNN architectures to balance performance with sustainability, exploring new algorithms and energy-efficient methods.

We hope this work inspires further efforts to develop eco-friendly AI technologies that align with global sustainability goals.

\section*{Acknowledgment}
This work was partially supported by projects FAIR (PE0000013) and SERICS (PE00000014) under the MUR National Recovery and Resilience Plan funded by the European Union - NextGenerationEU and by PRIN 2020 project n.2020TA3K9N "LEGO.AI". Supported by the EC H2020RIA project “SoBigData++” (871042), PNRR MUR project  IR0000013-SoBigData.it.

\bibliographystyle{ACM-Reference-Format}
\bibliography{biblio}
\end{document}